# Theoretical framework for real time sub-micron depth monitoring using quantum inline coherent imaging

ALEXANDER WAINWRIGHT* AND KHALED MADHOUN

*The Department of Physics, University of Toronto, 60 St. George Street, Toronto, Canada M5S 3H6*
**Alexander.Wainwright@mail.utoronto.ca*

**Abstract:** Inline Coherent Imaging (ICI) is a reliable method for real-time monitoring of various laser processes, including keyhole welding, additive manufacturing, and micromachining. However, the axial resolution is limited to greater than 2 μm making ICI unsuitable for monitoring submicron processes. Advancements in Quantum Optical Coherence Tomography (QOCT), which uses a Hong-Ou-Mandel (HOM) interferometer, has the potential to address this issue by achieving better than 1 μm depth resolution. While time-resolved QOCT is slow, Fourier domain QOCT (FD-QOCT) overcomes this limitation, enabling submicron scale real-time process monitoring. Here we review the fundamentals of FD-QOCT and QOCT and propose a Quantum Inline Coherent Imaging system based on FD-QOCT. Using frequency entangled sources available today the system has a theoretical resolution of 0.17 μm, making it suitable for submicron real-time process monitoring.

## 1. Introduction

Inline coherent imaging (ICI) is an effective method of monitoring laser processes in real-time to improve the reliability and repeatability of various laser processes, including laser keyhole welding, additive manufacturing, laser micromachining, and laser surgery [1–5]. The system is based on a Michelson interferometer, similar to those used in optical coherence tomography (OCT). Unfortunately, most systems of this type are limited to an axial resolution greater than 2 μm, which limits the variety of applications the technique can monitor. For example, single or dual pulse femtosecond (fs) laser micromachining can modify material with a precision better than 1 μm, which is below the resolution of traditional ICI or OCT systems [6,7]. As a result, these ultrafine machining applications cannot be monitored in real-time, leading to poor consistency due to slight changes in the machining environment [8,9]. Fortunately, advances in Quantum Optical Coherence Tomography (QOCT), which uses a Hong-Ou-Mandel (HOM) Interferometer to replace the Michelson interferometer, may solve this problem. By using frequency entangled photons, HOM interferometry can offer a factor of 2 improvement in depth resolution when compared to a Michelson interferometer with the same spectral bandwidth and central wavelength [10–13]. Unfortunately, time-resolved QOCT is an extremely slow process. However, Fourier domain QOCT (FD-QOCT) overcomes this problem allowing for submicron scale real-time process monitoring.

    Here we will review the fundamentals of QOCT and FD-OCT before presenting the theoretical framework of a "quantum inline coherent imaging" (QICI) method based on the principles of FD-QOCT. Section 2 will first cover the theoretical framework for FD-QOCT. Section 3 will follow with the proposed setup and simulated results based on recent advancements in chirped frequency entangled sources to determine a theoretical resolution limit

available with today's technology. In Section 4, we discuss how such a system could provide advantages over the standard ICI system.

## 2. Theory

### Inline coherent imaging

ICI uses a Michelson interferometer to monitor manufacturing processes in real-time. The technique is based on Spectral domain OCT (SD-OCT), also known as Fourier domain OCT (FD-OCT), where measurements are taken coaxially with the processing laser beam. ICI is well suited for applications which require precise measurements at high acquisition rates. A schematic diagram of ICI is shown in Fig. 1.

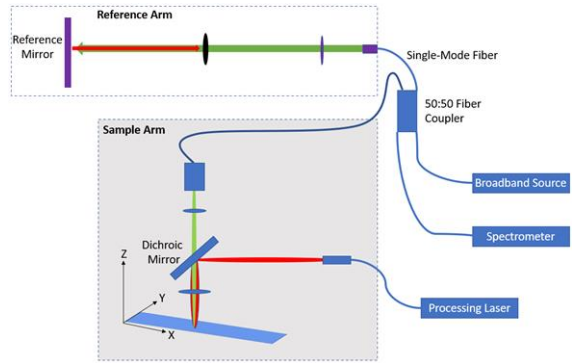

Figure 1: Schematic diagram of a standard ICI system used to monitor laser machining processes.

The OCT equation, derived in references [3] and [18], is used in ICI to understand the relationship between the difference in path length of light and the signal observed by the detector. The critical result for a gaussian beam profile, SD-OCT system is Eq. (1), which represents the intensity recorded at the detector for an individual A scan [3]:

$$|I(z)| = \frac{I_0}{2\sqrt{\pi}} \Delta k \left[ \sum_{i=1}^{N} A_i A_r \exp\left(-\frac{\Delta k^2}{4}(z \pm \Delta z_{i,r})^2\right) + \sum_{i=1}^{N-1} \sum_{j=i+1}^{n} A_i A_j \exp\left(-\frac{\Delta k^2}{4}(z \pm \Delta z_{i,j})^2\right) \right] \quad (1)$$

Where $I_0$ is the input beam intensity, $A_i$ is the reflectivity of the $i^{th}$ surface, $A_r$ is the reflectivity from the reference arm, $\Delta z_{i,r}$ is the displacement from the reference arm length to the $i^{th}$ surface, $z$ is the path length of the sample arm, and $\Delta z_{i,j}$ is the distance between the $i^{th}$ and $j^{th}$ surface. From the change in intensity as a function of z, it is possible to measure the location of each interface in a ICI measurement. ICI is presently limited to applications that need a depth resolution of greater than 1 micron. The theoretical limit of axial resolution ($\Delta z$) of an OCT system is half the coherence length, which is given by:

$$\Delta z = \frac{2 \ln(2)}{\pi} \frac{\lambda_0^2}{\Delta \lambda} \quad (2)$$

Where $\lambda_0$ is the central wavelength, and $\Delta\lambda$ is the spectral bandwidth measured as full-width half max (FWHM). For standard values used in ICI (a bandwidth of 32 nm and a central wavelength of 840 nm), the theoretical limit for depth resolution is 9.7 μm [19,20]. To achieve better resolutions, a broader spectrum and shorter wavelength are required. However, OCT systems constructed in the visible and near ultraviolet spectrum have not been able to achieve sub-micron depth resolutions [17]. It should be noted that OCT using a laser plasma source of soft x-rays and extreme ultraviolet light has achieved nm scale depth resolution [18]. However, these systems have a very low potential for industrial applications in their present form [22]. A challenge with state of the art ICI systems is a low imaging rate when imaging narrow cavities [3,15]. The low imaging rate is caused by light loss (up to 99%) from light scattering inside the cavities formed by processes like laser ablation and laser keyhole welding [3].

Quantum optical coherence tomography:

A comprehensive grasp of the underlying interferometric system is necessary to understand QOCT. The HOM interferometer has two key components, a 50:50 beam splitter and entangled photon pairs generated through the nonlinear process of spontaneous parametric down-conversion. Although a HOM interferometer is not an interferometer in the classical sense, we can derive the interference effect produced by the system. We start our derivation by considering the quantum mechanical beam splitter, as shown in Fig. 2, which is based on Ref. [19].

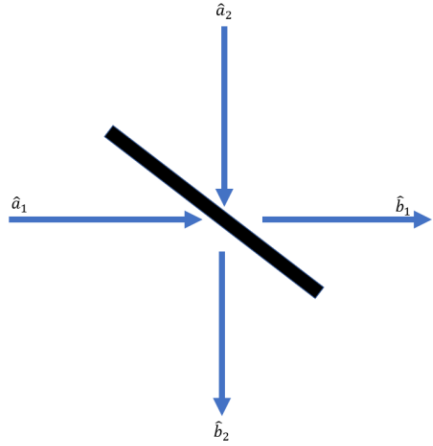

Figure 2: Diagram of the quantum mechanical beam splitter.

As derived in several sources, the outputs from the beam splitter ($\hat{b}_1$ and $\hat{b}_2$) can be written in terms of the transmittance and reflectance of the beam splitter [19,20].

$$\hat{b}_1 = \sqrt{T}\hat{a}_1 + \sqrt{R}\hat{a}_2 \qquad (3)$$

$$\hat{b}_2 = \sqrt{T}\hat{a}_2 - \sqrt{R}\hat{a}_1 \qquad (4)$$

Where $T$ is the transmittance and $R$ is the reflectivity of the beam splitter. For an input of wavefunction, $|\psi_{in}\rangle = |1_{a_1}, 1_{a_2}\rangle$, representing photons entering from both $\hat{a}_1$ and $\hat{a}_2$, the output wavefunction can be written as:

$$|\psi_{out}\rangle = (T-R)|1_{b_1}, 1_{b_2}\rangle + \sqrt{2TR}\left(|2_{b_1}, 0_{b_2}\rangle + |0_{b_1}, 2_{b_2}\rangle\right) \qquad (5)$$

For a 50:50 beam splitter, which is the most common type of beam splitter used in HOM interferometry, the transmittance and reflectance are equal. In this case, the first term of Eq. 5 goes to zero, resulting in an output wavefunction of:

$$|\psi_{out,50:50}\rangle = \sqrt{0.5}\left(|2_{b_1}, 0_{b_2}\rangle + |0_{b_1}, 2_{b_2}\rangle\right) \qquad (6)$$

The output of the beam splitter will be important to our understanding of the HOM interferometer.

Hong-Ou-Mandel interferometer

After discussing the quantum mechanical beam splitter, the next step in understanding the HOM interferometer is the entangled photons produced by spectral parametric down conversion (SPDC). It can be shown that the resulting wavefunction from SPDC is given by [25,26]:

$$|\psi\rangle = |0_s, 0_i\rangle + \int dv_s dv_i \, \emptyset(v_s, v_i) \hat{a}_s^\dagger(v_1), \hat{a}_i^\dagger(v_2)|0\rangle \tag{7}$$

Where the subscript "s" indicates the signal photon and the subscript "i" indicates the idler photon as shown in Fig. 3.

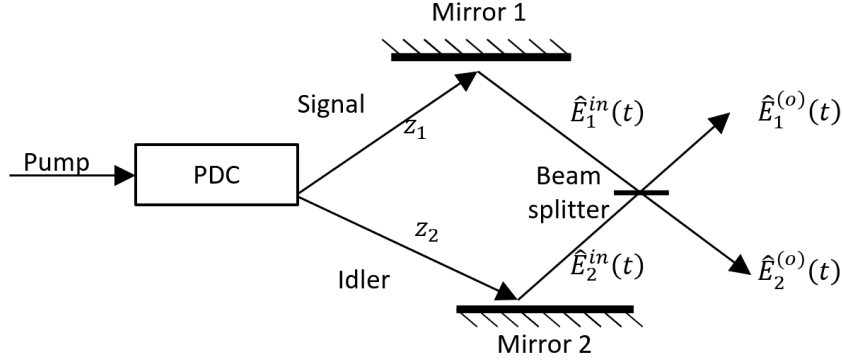

Figure 3: Hong-Ou-Mandel Interferometer diagram.

For a single pumping frequency at $2\omega_0$, the joint spectrum amplitude (JSA) can be written as:

$$\emptyset(v_s, v_i) = V_p(r, t)\delta(v_s + v_i - 2\omega_0)\psi(v_s) \tag{8}$$

With the pump field function $V_p(r, t)$ given by:

$$V_p(r, t) = \frac{1}{\sqrt{2\pi}} \int dv_p \alpha_p(v_p) \exp\left(i(k_p * r - v_p t)\right) \tag{9}$$

Where $\alpha_p(v_p)$ is the pump spectral amplitude. The wave function of the pumping field is then proportional to:

$$\psi(v_s) \sim \exp\left(-\frac{iL\Delta k}{2}\right) \text{sinc}\left(\frac{L\Delta k}{2}\right) \tag{10}$$

Note that $\Delta k$ is dependent on the type of phase matching used [23,26]. For a broadband pumping source, we note that the JSA takes the form of:

$$\emptyset(v_s, v_i) \sim \alpha_p(v_i + v_s)\psi(v_s) \tag{11}$$

Following the derivation of [19], it is then possible to solve for the output of the system using an input electric field of:

$$\hat{E}_m^{in}(t) = \frac{1}{\sqrt{2\pi}} \int dv \, \hat{a}_m(v) \exp\left(-iv\left(t - \frac{z_m}{c}\right)\right) \tag{12}$$

Where $m \in [s, i]$, $z_m$ is the optical path for the down-converted photons. We then find that the output of the electric fields from the beam splitter is given by:

$$\hat{E}_1^{(o)}(t) = \sqrt{T}\hat{E}_s^{(in)}(t) + \sqrt{R}\hat{E}_i^{(in)}(t) \tag{13}$$

$$\hat{E}_2^{(o)}(t) = \sqrt{T}\hat{E}_i^{(in)}(t) + \sqrt{R}\hat{E}_s^{(in)}(t) \tag{14}$$

The probability of detection is then given by the correlation function:

$$G^2(t_1, t_2) = \langle \hat{E}_1^{(0)\dagger}(t_1)\hat{E}_2^{(0)\dagger}(t_2)\hat{E}_2^{(0)}(t_2)\hat{E}_1^{(0)}(t_1)\rangle \tag{15}$$

After a lengthy calculation which is shown fully in reference [19], the coincidence count ($N_c$) is found to be proportional to the time average of the two-photon correlation function, which goes as:

$$N_c \sim \int_{-\infty}^{\infty} dt_1 dt_2 G^2(t_1, t_2) \sim 1 - \frac{2TR}{T^2 R^2} exp\left(-\frac{(\sigma \Delta z)^2}{2c^2}\right) \quad (16)$$

A full derivation of this result can be found in [23]. In Eq. 16, $\sigma$ is the spectral width of the photon wavefunction, and $\Delta z$ is the difference in the path length of the idler and the signal photons. By varying the length of the reference arm, $\Delta z$ can be changed to observe the HOM dip. Note that the minimum value of the HOM dip corresponds to the point where the reference arm and sample arm are the same length. By knowing the length of the reference arm at any given time, it is possible to perform an interferometry experiment in the time domain. This process is the foundation of time domain OCT (TD-QOCT) and is very slow as the reference mirror needs to be moved to acquire each data point. Depending on the number of data points used, TD-QOCT can take minutes or many hours, making this process unsuitable for applications requiring real time feedback. However, with all things being equal Fourier domain QOCT (FD-QOCT) can be completed in seconds [13,24,25].

FD-QOCT:

The easiest way to derive the results of FD-QOCT is to use the joint spectrum approach as done in [25]. In this approach, we consider the design of FD-QOCT as shown in Fig. Figure 4. Where after SPDC in a nonlinear crystal, one photon propagates in the object placed in the sample arm of the interferometer and the other one is reflected from a mirror in the reference arm. Both photons overlap at the beam splitter and the two spectrometers take a measurement of the wavelength-dependent coincidence of their simultaneous detection, which is known as the joint spectrum intensity (JSI) ($C(v_s, v_i)$).

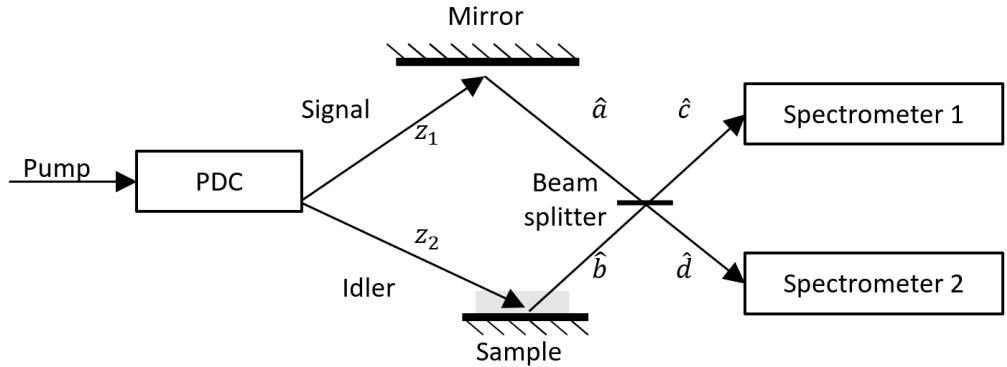

Figure 4: Schematic diagram of QOCT. The reference arm is represented by path $z_1$ and the object arm is represented by path $z_2$.

The JSI of the system can be found by considering the JSA of the two photons defined by:

$$\emptyset(v_1, v_2) = \exp\left(\frac{(v_s - v_i)^2}{2\sigma_d^2} - \frac{(v_s + v_i)^2}{2\sigma_p^2}\right) \quad (17)$$

The JSI is defined by $|\emptyset(v_s, v_i)|^2$, where $\sigma_p$ is the spectral width of the pump light and $\sigma_d$ is the spectrum width determined by the SPDC crystal and the spectral mode characteristics. For an ideal continuous wave laser, the spectrum produced is a single frequency resulting in $\sigma_p \approx 0$. However, for a broad band spectrum, it is possible to have more information carried per pulse allowing for algorithmic near-perfect dispersion cancellation and artefact removal [26].

When the two photons generated in the SPDC propagate along the paths marked "a" (the sample arm), "b" (the reference arm),"c", and "d" in Fig. 4, the electric fields can be represented in terms of raising operators as done in section 2.2, ($\hat{a}^\dagger(v), \hat{b}^\dagger(v), \hat{c}^\dagger(v), \hat{d}^\dagger(v)$). Just before entering the dispersive object being analyzed by the QOCT system, the wave function for the two photons can be written as:

$$|\psi\rangle = |0_s, 0_i\rangle + \int dv_s dv_i \, \emptyset(v_s, v_i) \hat{a}^\dagger(v_s), \hat{b}^\dagger(v_i)|0\rangle \tag{18}$$

While propagating on the sample arm, the photon wavefunction will acquire a phase which can be modeled as a transfer function ($f(v)$). In the reference arm, the photon is reflected from a reference mirror and experiences some temporal delay. The two-phase contributions modify the two-photon wave function to become:

$$|\psi(\tau)\rangle = \int dv_s dv_i \, \emptyset(v_s, v_i) f(v_s) e^{iv_i\tau} \, \hat{a}^\dagger(v_s), \hat{b}^\dagger(v_i)|0\rangle \tag{19}$$

Next the beams interact with the beam splitter which gives the final wavefunction:

$$|\psi(\tau)\rangle = \int dv_s dv_i \, \emptyset(v_s, v_i) f(v_s)[\hat{c}^\dagger(v_s) + \hat{d}^\dagger(v_s)][\hat{c}^\dagger(v_i) - \hat{d}^\dagger(v_i)]|0\rangle \tag{20}$$

Using this result to solve for the wavefunction that would represent the coincident photons, the only terms in Eq. 20 that are of interest are a product of $\hat{c}^\dagger$ and $\hat{d}^\dagger$:

$$|\psi(\tau)\rangle_{coinc} = \int dv_s dv_i \, \emptyset(v_s, v_i) \hat{c}^\dagger(v_i) \hat{d}^\dagger(v_s)[f(v_s)e^{iv_i\tau} - f(v_i)e^{iv_s\tau}]|0\rangle \tag{21}$$

We can then solve for the coincidence probability density:

$$p(v_s, v_i, \tau) = |\phi(v_s, v_i)|^2 \left| f(v_s)e^{iv_i\tau} - f(v_i)e^{iv_s\tau} \right|^2 \tag{22}$$

Note that the result of Eq. 22 is general for the time and frequency resolved forms of QOCT. For the frequency resolved case, the time delay ($\tau$) is fixed, and the reference mirror is stationary. The measurement signal of the two-dimensional modulated joint spectrum is then given by:

$$C = |\phi(v_s, v_i)|^2 M(v_s, v_i) \tag{23}$$

Where we define the modulation component as:

$$M(v_s, v_i) = |f(v_s)|^2 + |f(v_i)|^2 + 2Re(f(v_s)f^*(v_i)) \tag{24}$$

For a single-layer object, we can then derive an expression for $M(v_s, v_i)$. If an object consists of one dispersive layer of thickness $z_2$ placed at a distance $z_1$ from the zero optical path difference point, the total phase acquired in such an object can be expressed as:

$$f(v) = R_1 \exp(iz_1(B_0 + B_1 v)) + R_2 \exp\left(i(z_1(B_0 + B_1 v) + Z_2(B_0 + B_2 v + B_{22} v^2))\right) \tag{25}$$

Where $B_0$ is the wavenumber of light in air, $B_1$ is the inverse of the group velocity of light in air, $B_2$ is the inverse of the group velocity of light propagating through the dispersive layer and $B_{22}$ is the group velocity dispersion coefficient of the dispersive layer. $R_1$ and $R_2$ are the reflectivity of the layer interfaces. Substituting this result into the modulation component, Eq. 23 becomes:

$$\begin{aligned}
\frac{1}{2} M(v_s, v_i) = & -R_1^2 \cos[2z_1 B_1 (v_s - v_i)] \\
& -R_2^2 \cos[2z_1 B_1 (v_s - v_i) + 2z_2(B_2 + B_{22}(v_s + v_i))(v_s - v_i)] \\
& +R_1 R_2 \cos[2z_2(B_0 + B_2 v_s + B_{22} v_s^2)] \\
& +R_1 R_2 \cos\left(2z_2(B_0 + B_2 v_i + B_{22} v_i^2)\right) \\
& -R_1 R_2 \cos[2z_1(v_s - v_i) + 2z_2(B_0 + B_2 v_s + B_{22} v_s^2)] \\
& -R_1 R_2 \cos[2z_1(v_s - v_i) + 2z_2(B_0 + B_2 v_i + B_{22} v_i^2)] \\
& +(R_1^2 + R_2^2)
\end{aligned} \tag{26}$$

Each term in Eq. 26 gives different information about the system. The first two terms convey information about the object's structure, while the next two contribute to the stationary artefact peak placed at a distance from zero optical path difference. The last two terms correspond to a stationary artefact peak that is always positioned midway between the two interfaces. Because the location of these artifact peaks is known, they can easily be removed post processing. An example output of this system can be seen in Fig. 5.

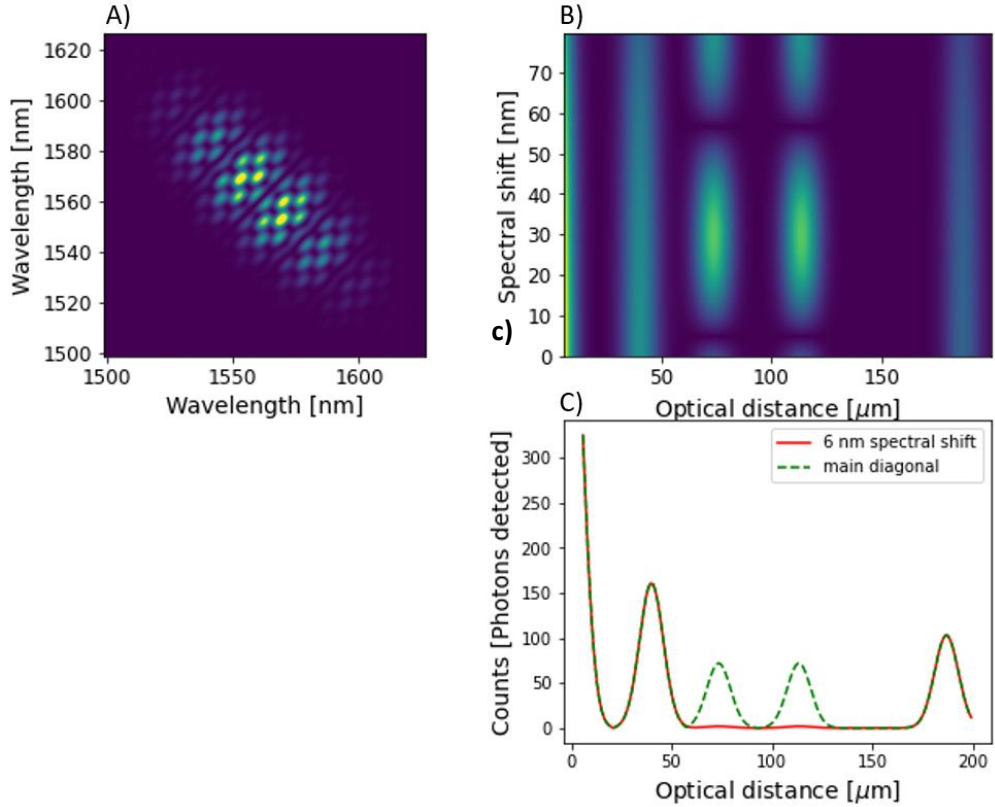

Figure 5: Plot illustrating the process of FD-QOCT for a double layer sample consisting of 100 μm thick quartz positioned 40 μm away from zero optical path difference. The reflectivity of the first surface was set to 50%, and the second surface was set to 40%. Plots were produced following the procedure outlined in Ref. [25]. (a) JSI corresponding to a double-layer object: (b) Fourier transforms of the JSI. Note that the intensity of each artefact peak drops to zero for different spectral shifts. The optical distance axis is the difference in propagation time multiplied by the speed of light in a vacuum. As a result, the distance estimated for the sample's thickness is off by a factor of the index of refraction, 1.47 for quartz. (c) is the Fourier transform of the main diagonal of the HSI and the 6 nm spectral shift. Each true peak is associated with one artifact peak across the entire spectrum, and the number of counts of the main peak is constant across the spectra.

To convert from the spectral domain into the time domain, a Fourier transform of the two-dimensional modulated joint spectrum is taken. To convert to a depth profile, the results in the time domain are multiplied by the speed of light in each medium. Note that the diagonal spectra which corresponds to when $v_i = v_s$, will have information on the depth profiles as well as contain artificial peaks. In the off-diagonal elements, which correspond to $v_i = v_s + \Delta v$, the system does not contain any artifact peaks. As a result, by comparing diagonal and off diagonal peaks it is possible to suppress artifacts during processing and resolve the depth profile of a structure [25]. There are also other algorithmic approaches to reducing artifacts and almost complete dispersion cancelation, which have been suggested in Refs. [23], [26], and [27]. The entire FD-QOCT process requires only one scan to be conducted which significantly reduces the time required for QOCT [21,28].

## 3. Proposed system

Here we aim to apply the same innovation of FD-QOCT to ICI. The system designed is based around the preexisting Michelson interferometer ICI, where the Michelson interferometer is replaced with a HOM interferometer as shown in Fig. 6. Several different broadband entangled sources were simulated to determine the optimum design based on the present literature.

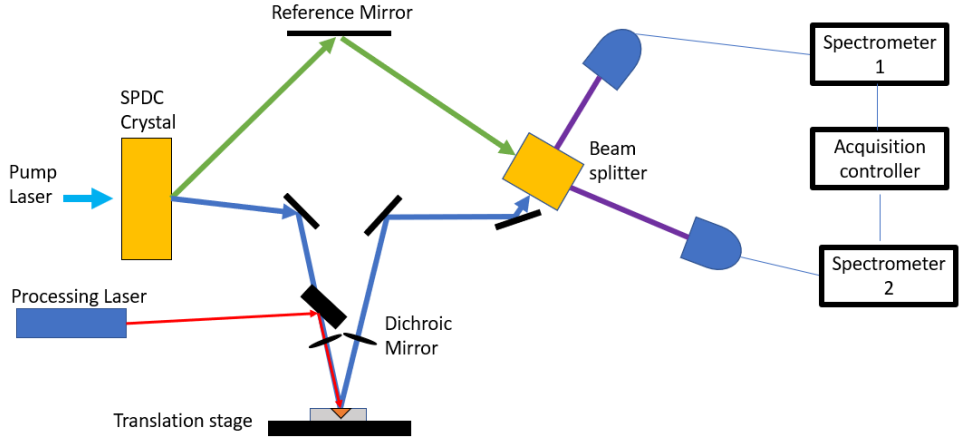

Figure 6: Preposed design of QICI using a HOM interferometry system.

Simulations:

We can derive our M function in an analogous manner to that used in the case of the single layer FD-QOCT, with $z_2$ and $R_2$ set to zero. Using these inputs in Eq. 26, we find:

$$\frac{1}{2}M(v_s, v_i) = -R_1^2 \cos[2z_1 B_1 (v_s - v_i)] + R_1^2 \qquad (27)$$

To understand how the system would behave in a real-world application, the system was simulated with very low reflectivity. From a keyhole weld, only a small portion of the light will be reflected to the interferometer. According to Galbraith's estimation, approximately 1% of the input signal is recovered from a keyhole weld because of the scattering that occurs at the bottom of the weld as discussed in Ref. [3]. This can be modeled as $R_1 = 0.01$. To simulate real world spectrometer data, a spectral resolution of 0.1 nm/pixel was used. Although ultrahigh resolution spectrometers can have near an order of magnitude better spectra resolution, simulating with a standard 0.1 nm/pixel resolution will give a better idea of the real-world performance that can be found using this system when other sources of loss are considered.

Results:

Like standard ICI, the depth resolution is dependent on both the central wavelength and the bandwidth of the entangled photon source. Our modeling of the QICI system, using equations 23 and 27 indicate that the optimum depth resolution is achieved at short central wavelengths and for broad bandwidths, as shown in Figure 7. Note that we define the depth resolution as the FWHM of the counts profile.

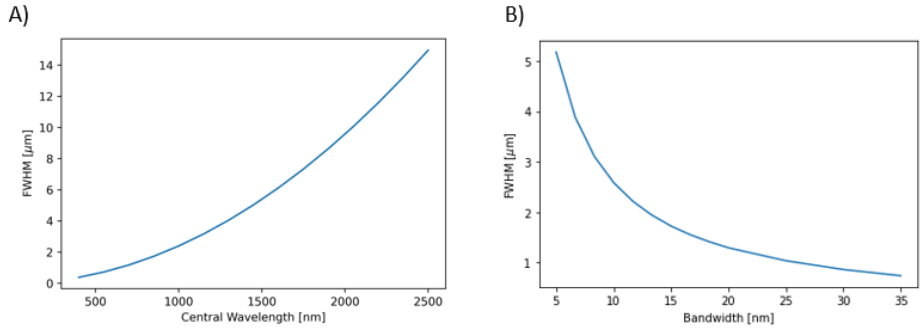

Figure 7: A) Depth resolution defined in terms of FWHM as a function of the central wavelength for a 100 nm spectral bandwidth, and a depth measurement of 100 µm. B) Depth resolution as a function of the bandwidth of the entangled photon source for a central wavelength of 600 nm.

A plot of the interferometric normalized count profiles for the sources in Table 1 is shown in Fig. 8. The depth resolution is found to be independent of the depth measured, as expected.

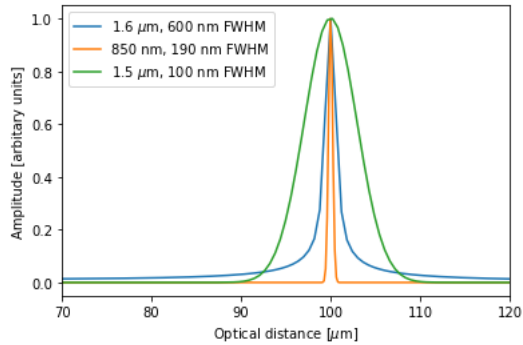

Figure 8: Plot of the scan output of three different sources measuring a depth of 50 µm. Note that the FWHM is dependent on the spectrum central wavelength and the FWHM of the entangled source.

Table 1: Depth resolution for each of the tested spectra.

| Broadband spectra tested | | | |
| --- | --- | --- | --- |
| Source | Reference [31] | Reference [12] | Used by Reference [25] |
| Central wavelength | 1.6 µm | 850 nm | 1.5 µm |
| Bandwidth (FWHM) | 600 nm | 190 nm | 100 nm |
| Depth resolution | 0.61 µm | 0.17 µm | 5.8 µm |

Based on the data collected, the optimum source would be the 850 nm central wavelength with a 190 nm FWHM spectral width, giving the best depth resolution of 0.17 µm.

## 4. Discussion

To date, the best reported depth resolution for standard inline coherent imaging is 5 µm [30], while the best reported standard OCT has a depth resolution of 3 µm [32,33]. The results collected in the simulations clearly indicate that an order of magnitude improvement in depth resolution of real time process monitoring is achievable with the basic QICI system presented. When comparing the performance of temporal domain QOCT systems, the best reported depth resolution is 0.54 µm [12]. Using the same parameters as this experiment but in the spectral domain, a resolution of 0.17 µm is predicted. According to the analytical investigation conducted in ref. [26], the time domain HOM system will have a lower fundamental limit in the depth resolution than the Fourier domain measurement by a factor of $\sqrt{3}$. This result indicates that there is significant room for improvement in the depth resolution of both TD and FD QOCT systems.

A depth resolution of 0.17 μm would allow for the monitoring of sub-micron scale machining processes. Which could include sub-cellular biopsies, nanofabrication, and microfluidic applications. One application of interest is double pulse fs-laser processing which has extremely precise material removal [6,33]. For these processes to achieve wide adoption, a system like QICI is required to monitor the system in real time [34].

One possible limiting factor for implementing a QICI system is the spectral resolution of the measurement systems. In the simulations, a spectra resolution of 0.1 nm was used as this is close to the standard resolution of spectrometers available on the market today. With the development of ultrahigh resolution spectrometers, with a resolution better than 0.01 nm per pixel, the spectral resolution is not expected to limit the implementation of QICI [34].

The use of a HOM interferometer has some significant advantages compared to other quantum mechanical interferometers. As discussed in Ref. [35], the HOM interferometer is free from frequency-independent noise unlike the Mach-Zehnder interferometer. However, the HOM interferometer is susceptible to fluctuations in the path length that may arise from vibrations or heating in the system [35]. This type of noise is common to both the QOCT and FD-OCT and has already been addressed sufficiently for QOCT to achieve sub μm scale depth resolution [12]. As a result, it is expected that signal to noise ratios of FD-OCT and QICI systems should be sufficient for imaging sub μm scale depth resolutions.

The signal loss experienced in conventional ICI was not observed in the low reflection simulations conducted for QICI. However, these simulations only provide a small glimpse into the complicated signal loss associated with ICI of Keyhole welds. One method being tested to improve the signal to noise ratio of ICI is the self-witnessing approach, where two reference arms are used to find a signal which would otherwise be lost in the background noise of the detectors [15]. It is feasible that such a system could also be implemented on a QICI setup if the signal lost in the weld is found to be a significant issue.

Clearly, QICI and QOCT systems have significant benefits over their classical counterparts with improved depth resolution. Using a spectral domain HOM interferometer, it is possible to overcome the historic challenge of slow QOCT systems, to allow for real time process monitoring. Here, we have used a simulation to demonstrate the improved depth resolution, as low as 0.17 μm, indicating that QICI has real promise. The next step in making QICI a reality is to construct such a system and explore its experimental potential.

**Funding.** NSERC Discovery grant RGPIN-2019-06518.

**Acknowledgments.** A. Wainwright would like to thank R.J.D. Miller and D.F.V. James for their support and mentorship in creating this work.

**Disclosures.** The authors declare no conflicts of interest.

**Data availability.** Data and codes underlying the results presented in this paper are available upon request from the corresponding author.